\documentclass[conference]{IEEEtran}
\IEEEoverridecommandlockouts
\usepackage{cite}
\usepackage{amsmath,amssymb,amsfonts}
\usepackage{algorithmic}
\usepackage{graphicx}
\usepackage{textcomp}
\usepackage{float}
\usepackage{xcolor}
\usepackage{booktabs}
\usepackage{multirow}
\usepackage{hyperref}
\usepackage{soul}

\def\BibTeX{{\rm B\kern-.05em{\sc i\kern-.025em b}\kern-.08em
    T\kern-.1667em\lower.7ex\hbox{E}\kern-.125emX}}

\usepackage{amssymb}
\usepackage{pifont}
\newcommand{\cmark}{\ding{51}}%
\newcommand{\xmark}{\ding{55}}%

\begin{document}

\title{Fusing Audio and Metadata Embeddings \\ Improves Language-based Audio Retrieval}
\author{\IEEEauthorblockN{Paul Primus$^1$, Gerhard Widmer$^{1,2}$}
\IEEEauthorblockA{$^1$Institute of Computational Perception \\ 
$^2$LIT Artificial Intelligence Lab\\
Johannes Kepler University Linz, Austria}
}

\maketitle

\begin{abstract}
Matching raw audio signals with textual descriptions requires understanding the audio's content and the description's semantics and then drawing connections between the two modalities.
This paper investigates a hybrid retrieval system that utilizes audio metadata as an additional clue to understand the content of audio signals before matching them with textual queries. 
We experimented with metadata often attached to audio recordings, such as keywords and natural-language descriptions, and we investigated late and mid-level fusion strategies to merge audio and metadata. 
Our hybrid approach with keyword metadata and late fusion improved the retrieval performance over a content-based baseline by $2.36$ and $3.69$ pp. mAP@10 on the ClothoV2 and AudioCaps benchmarks, respectively. 
\end{abstract}


\section{Introduction}

Language-based audio retrieval systems search for audio recordings given a textual description of the desired content. Such textual queries enable low-effort retrieval as they permit users to intuitively express arbitrary concepts of interest such as acoustic events, temporal relationships, or sound quality.

\begin{figure}
    \centering
    \includegraphics{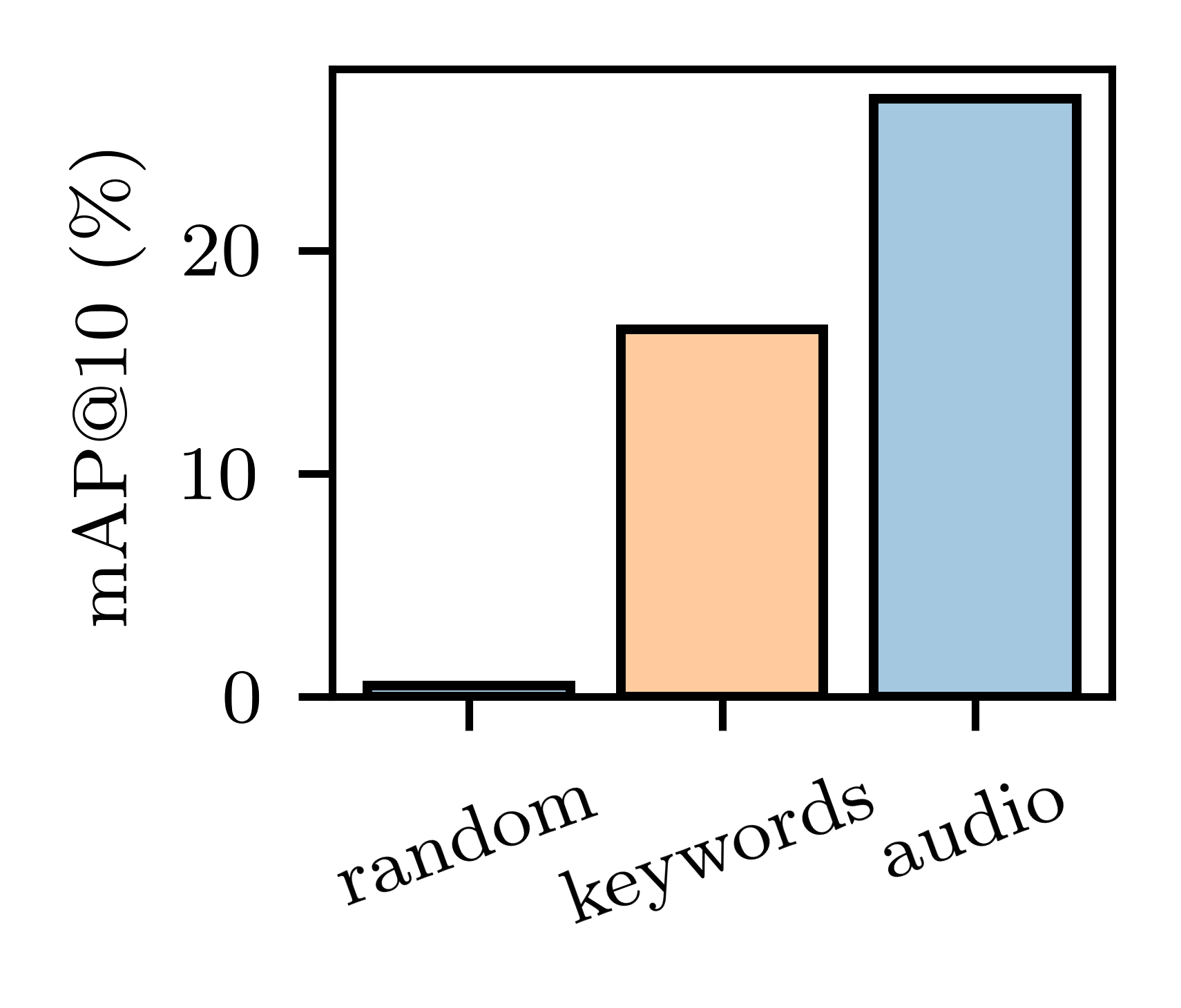}
    \includegraphics{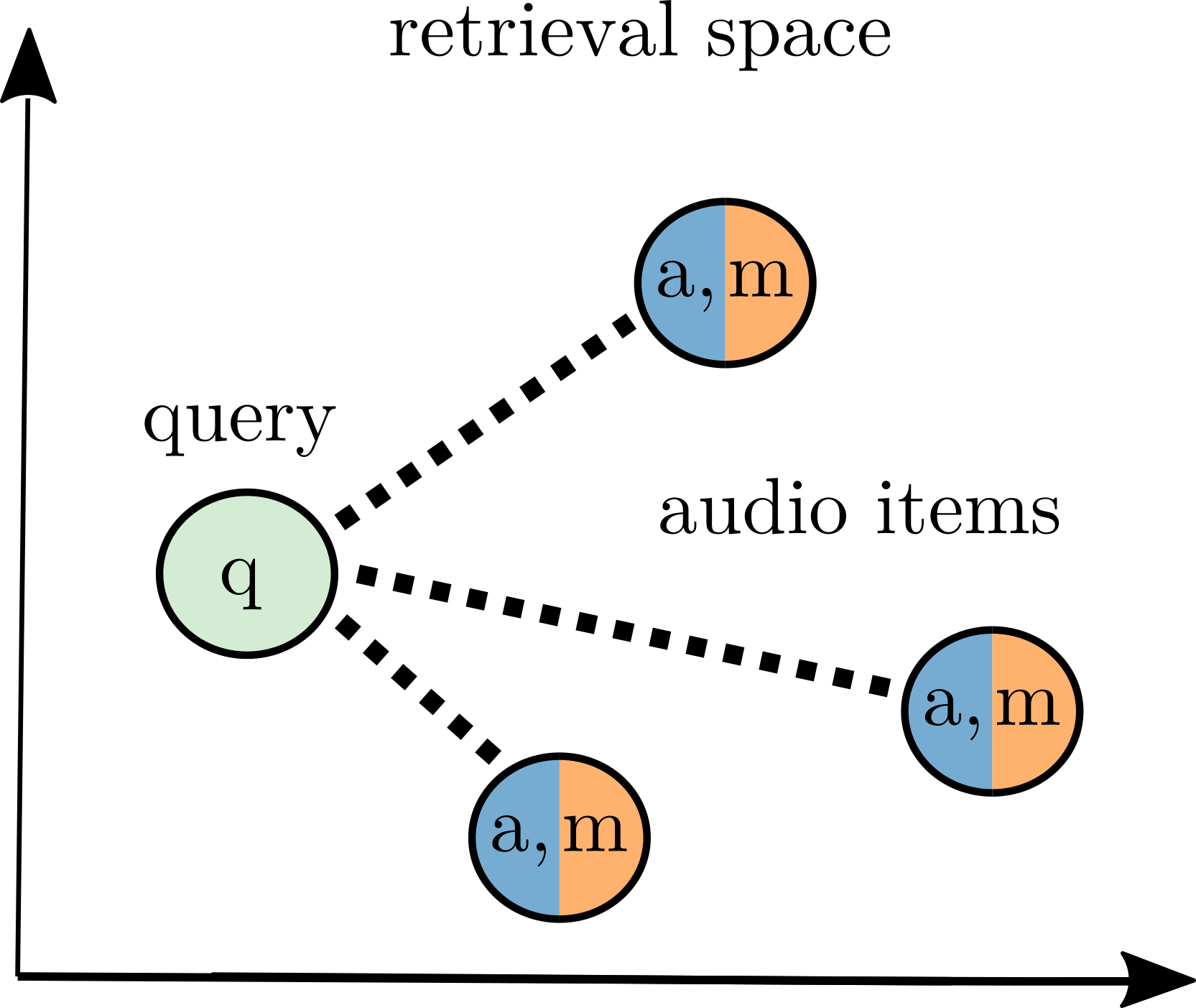}
    \caption{Left: Comparison of pure metadata- and content-based methods (orange and blue, respectively) on the ClothoV2 benchmark. Right: Illustration of the multimodal retrieval space of our hybrid approach. Audio signal (blue) and metadata (orange) are embedded and fused to represent an item $(a,m)$. The similarity to an embedded query $q$ (green) is measured via distance.}
    \label{fig:metadata_retrieval}
\end{figure}

However, matching the raw audio signals with textual queries is challenging. 
Audio retrieval systems are commonly based on the dual-encoder architecture that projects the query and the audio recordings into a shared multimodal metric space \cite{journal, pooling_and_objective,netrvlad,loss_comparison} (for another approach, see \cite{killing_birds}). This allows all the audio items to be ranked by their distance to the query. We will refer to this shared space as the \textit{retrieval space}. 
Previous works have explored multiple paths to improving natural language-based audio retrieval systems, such as using better pre-trained embedding models \cite{DCASE2023}, augmentation techniques for both modalities \cite{DCASE2022}, and artificial captions generated with large language models from metadata \cite{wavcaps,cacophony,DCASE2023}.
All of these previous works are based on content-based retrieval that derives the audio items' representation in the retrieval space exclusively from the audio signal.
However, additional information about the audio recording is often available in practice.
For example, \textit{FreeSound}\footnote{\url{https://freesound.org/}}, a popular public repository of Creative Commons licensed sounds, instructs the uploading users to specify a title and at least three keywords describing the audio recording.
Figure \ref{fig:metadata_retrieval} (left) demonstrates that a dual encoder retrieval system that, instead of the audio signals, simply embeds those keywords into the retrieval space performs significantly better than the random baseline. We argue that this additional metadata should be exploited for retrieval. 
In this work, we will thus explore whether content-based audio retrieval systems can be improved by using metadata in addition to the audio signal to match audio items to queries (Figure \ref{fig:metadata_retrieval} right). We will call systems that use both the audio recordings and their metadata \textit{hybrid} methods because they are a mixture of pure content-based and pure metadata-based retrieval systems. 


\section{Related Work}

Using metadata for language-based audio retrieval systems is not unheard of. Recent work \cite{wavcaps} generated artificial audio captions from metadata with the help of large language models. To this end, the authors used a variety of audio sources with diverse annotations, such as temporally strong and weak labels, open-set tags, or multi-sentence textual descriptions. They prompted ChatGPT to convert the metadata into a single-sentence description and used the newly created audio-text pairs for training. Similarly, \cite{cacophony} used a few-shot prompting approach with ChatGPT to convert descriptions into captions. Other recent work \cite{DCASE2023} used keywords associated with audio recordings and ChatGPT to augment audio captions. Altogether, these previous methods operate on textual inputs during training by converting metadata into artificial captions. This inflates the training set size but completely neglects the metadata during inference. In contrast, our hybrid retrieval method uses the available metadata as an additional piece of information to match audio recordings and queries during training and inference.

\section{Methodology} \label{sec:method}

We use two independent modality encoders to embed an audio signal and its corresponding metadata into separate embedding spaces. The resulting representations are then fused to obtain a single representation for an \textit{item} (i.e., an audio-metadata pair). This fused embedding is then projected into the shared \textit{retrieval space} where it is matched with embedded \textit{queries}.
For a search request, the textual query is also embedded into the shared retrieval space, and the $K$ closest items (measured via cosine similarity) are presented to the user. 

\subsection{Metadata}
\begin{figure}[t]
    \centering
    \includegraphics{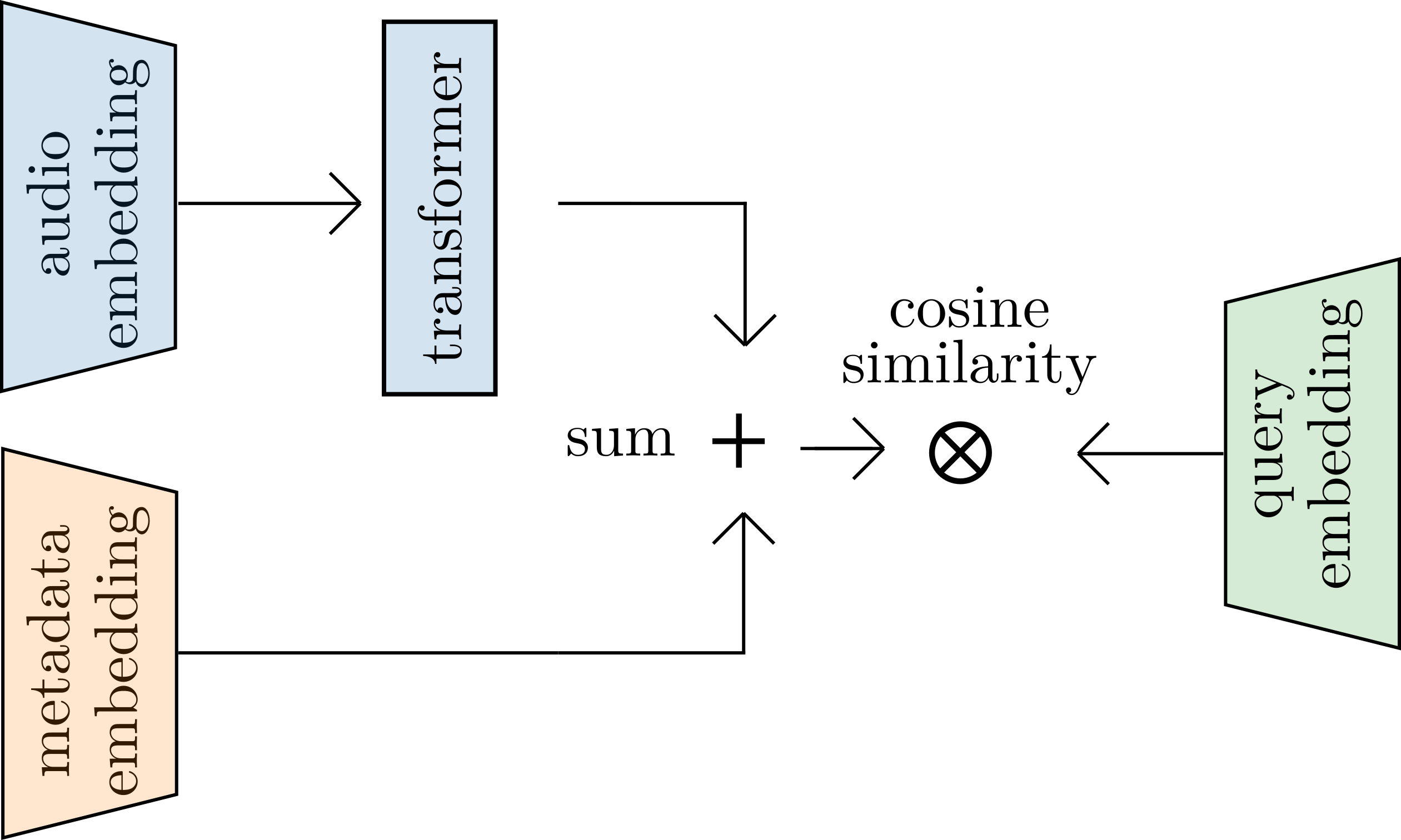}
    \caption{Late Fusion of audio (blue) and metadata (orange). The fused representation is matched with the embedded query (green) via cosine similarity.}
    \label{fig:latefusion}
\end{figure}
 
 Audio metadata comes in a variety of forms (structured or unstructured) and at different levels of temporal granularity (weak or strong labels).\footnote{By "temporally strong" and "weak" labels, we mean annotations with and without precise temporal event boundaries, respectively.} We restrict our investigation to keywords and natural language descriptions, as they are comparably cheap to collect and available for the two most popular audio-caption data sets, AudioCaps \cite{audiocaps} and ClothoV2 \cite{clotho}. In particular, we will consider three categories of metadata:
\begin{itemize}
    \item \textbf{Closed Set (CS) of Tags}: Temporally weak labels for a fixed number of acoustic events, such as descriptive tags chosen from a predefined list. 
    \item \textbf{Open Set (OS) of Tags}: Temporally weak labels not restricted to a fixed number of acoustic events, such as arbitrary descriptive keywords chosen by the user.
    \item \textbf{Full-Sentences (FS) Descriptions}: Single-sentence natural language descriptions such as descriptions used as captions for audio recordings.
\end{itemize}
\subsection{Audio, Metadata \& Query Embedding}


Both natural language queries and metadata can be represented as text. We, therefore, share a single text embedding model for query and metadata embedding and denote this model as $\phi_{t}$. To encode the CS and OS tags, we convert them to whitespace-separated lists of keywords. 

We further use a pretrained audio embedding model $\phi_a$ to compress the audio signals of varying lengths into sequences of embeddings. These varying-length sequences are then pooled into single vectors and projected into the retrieval space. Previous work \cite{pooling_and_objective, netrvlad,timing} has demonstrated that learnable pooling operations yield favorable retrieval results compared to simple non-parametric pooling operations like mean or max aggregation. We, therefore, use multiple transformer encoder layers \cite{attention} to convert the sequential output of the audio embedding model into a single vector embedding (similar to \cite{timing}). This is done by adding a fixed positional encoding to the audio encoder output and appending a global audio token to the sequence. This special token is initialized to the mean of the sequence plus a learnable bias. The whole sequence is passed through the transformed layers, and the transformed global audio token is used to represent the audio signal. 

\subsection{Audio-Metadata Fusion}
We experiment with two strategies to combine audio and metadata embeddings $\phi_a$ and $\phi_{t}$, respectively, to a single embedding model $\phi_f$. 

\textbf{Late fusion} (illustrated in Figure \ref{fig:latefusion}) is done by summing the output vectors of the audio and the metadata embedding models. This combination of the modalities is conceptually simple, but it does not allow for any crossmodal interactions between metadata and audio. 

\begin{figure}[t]
    \centering
    \includegraphics{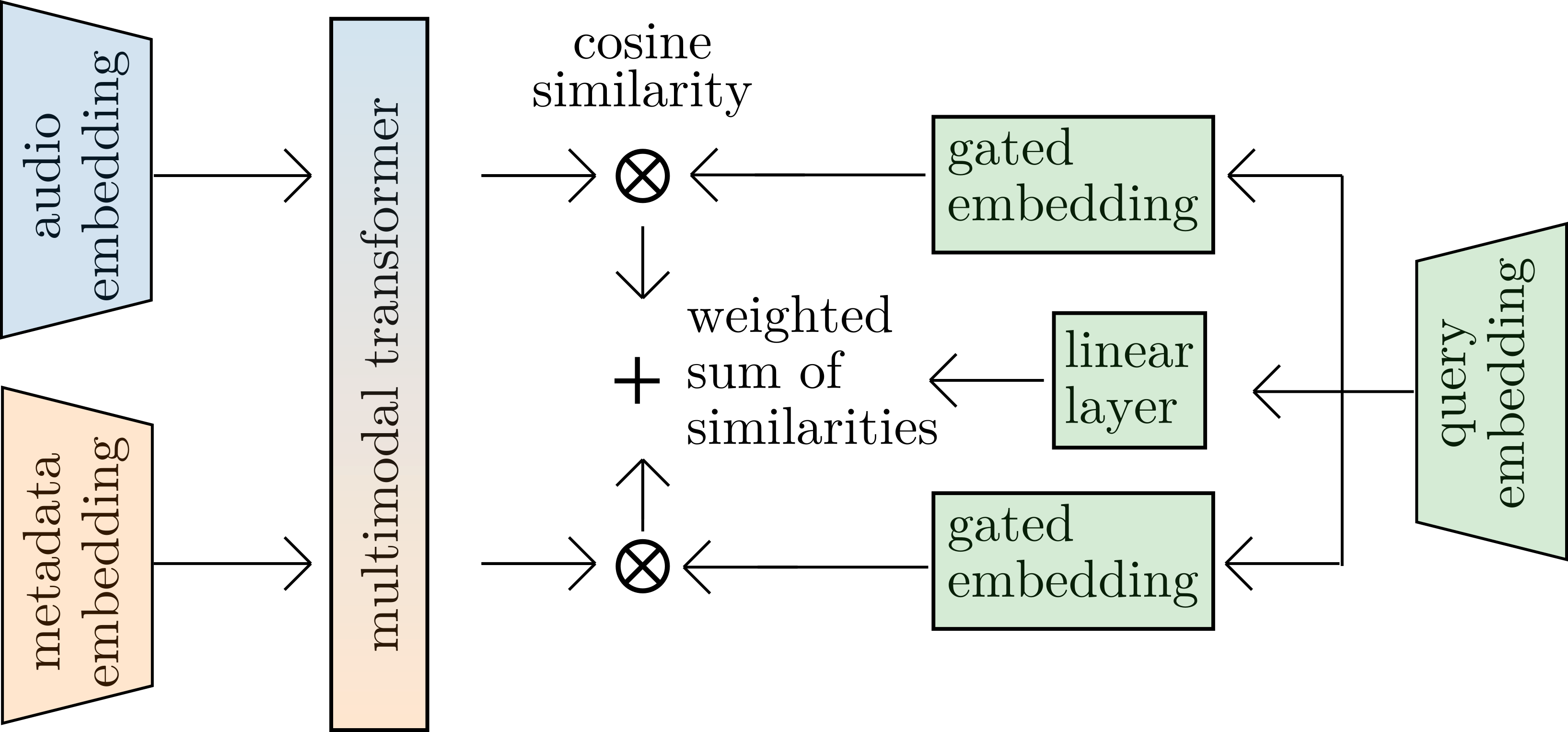}
    \caption{Mid-Level fusion: The matching of fused audio and metadata embeddings is inspired by the multimodal transformer \cite{MMT}.}
    \label{fig:midlevelfusion}
\end{figure}

\textbf{Mid-level fusion} (illustrated in Figure \ref{fig:midlevelfusion}), on the other hand, is more complex, but it allows interaction between the modalities. The architecture is motivated by the multimodal transformer (MMT) introduced for language-based video retrieval \cite{MMT}. The audio is processed as described in the previous section, but for mid-level fusion, the metadata embedding vector is appended to the audio embedding sequence, and the joint sequence is processed via multiple transformer layers. Two gated embedding modules \cite{gated_units} are used to convert the query embedding vector into an audio-query and a metadata-query vector, which are matched with the transformed global audio and metadata token, respectively. The resulting scores are then combined via weights derived from the query vector.

\section{Experimental Setup} \label{sec:exp}

\subsection{Datasets \& Benchmarks}

\begin{figure}[t]
    \centering
    \includegraphics{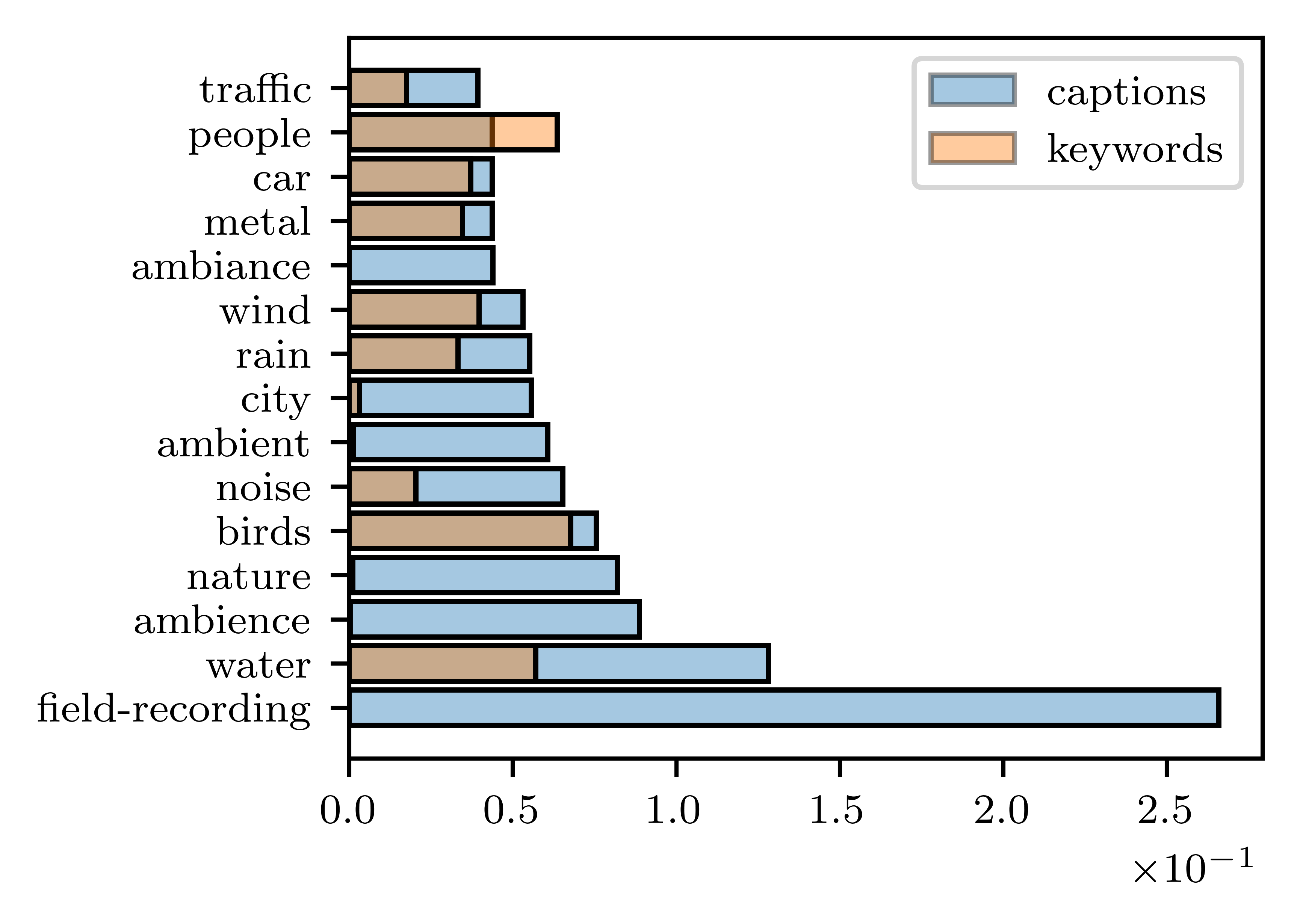}
    \caption{Relative frequency of the 15 most common keywords in ClothoV2 (blue) and their corresponding frequencies in the audio captions (orange).}
    \label{fig:word_distribution}
\end{figure}

We experimented with two audio-retrieval benchmark datasets:
ClothoV2 \cite{clotho} contains $15$-$30$ second audio recordings and captions that are between 8 and 20 words long. The provided training, validation, and test split contain 3840, 1045, and 1045 recordings, respectively; each recording is associated with five human-generated captions. Each audio recording also has a list of open-set keywords, which we will use as metadata. Figure \ref{fig:word_distribution} shows the frequencies of the 15 most common keywords and their corresponding frequency in the captions; it implies that keywords and queries overlap frequently. Furthermore, since each audio recording is associated with five distinct captions, we can simulate the availability of full-sentence descriptions as metadata. To that end, we use one of the captions as a query and another one as metadata during training and validation. It is important to stress that this setup simulates ideal conditions where the similarity between the query and the metadata is very high. 

AudioCaps \cite{audiocaps} consist of $51,308$ audio recordings taken from AudioSet \cite{audioset}. Each training and validation recording is associated with one and five human-written captions, respectively. The audio recordings' length is roughly 10 seconds, and the captions are, on average, 9.8 words long. Each audio is labeled with one or multiple acoustic event tags. An overview of the 527 classes used in AudioSet is given on the AudioSet website\footnote{\url{https://research.google.com/audioset/ontology/index.html}}. We will use those CS tags as metadata.

\subsection{Pretrained Models}

For audio embedding, we employed a pretrained efficient CNN model \cite{mn40_as} based on the MobileNetV3 \cite{mobilenet_v3} architecture (model ID: mn40\_as\_ext). The selected model was pre-trained on AudioSet \cite{audioset} using knowledge distillation from audio spectrogram transformers \cite{passt}. It achieves state-of-the-art performance on the AudioSet benchmark (mAP $48.7$) and other downstream tasks \cite{hear}. This architecture is particularly suitable for our tasks because it handles audio recordings of arbitrary length and returns a sequence of audio embeddings.

For description and metadata embedding, we used BERT (model ID: bert-base-uncased) \cite{bert}. The input text was pre-processed by transforming all characters to lowercase and removing punctuation. The resulting strings were tokenized with the WordPiece tokenizer \cite{wordpiece}, padded to the maximum sequence length in the current batch, and truncated if they were longer than 32 tokens. The transformed CLS token represents the compressed text.

\subsection{Optimization}

The modality encoders were jointly optimized using gradient descent and the NT-Xent \cite{NTxent} loss with a batch size of 32. We used the Adam update rule \cite{adam} for 25 epochs, with one warmup epoch. Thereafter, the learning rate was reduced from $2 \times 10^{-5}$ to $10^{-7}$ using a cosine schedule. The hyperparameters of the optimizer were set to PyTorch's \cite{pytorch} defaults. We further used SpecAugment \cite{specaugment} during training.

\subsection{Evaluation Metrics}

We evaluated the retrieval systems on the benchmarks' test item-query pairs. All results were averaged over three runs. We set the number of items to present to the user, K, to 10 and use the mean average precision at K, map@K, as our main comparison criterion. The map@K metric corresponds to a weighted average of the inverse ranks, with the weight being 1 if the correct item is among the top K results and 0 otherwise. In addition to that, we also report the recall among the top 1, 5, and 10 retrieved results. Unfortunately, the exact pairwise correspondence between $a_i$ and $q_j$ is not known for the case $i \neq j$, but it is common practice to assume that these pairs do not match. Consequently, the reported metrics are lower bounds for the actual performance; previous work has highlighted that the actual performance is likely higher \cite{journal}.

\section{Results \& Discussion}

\begin{table*}[ht]
\centering
\caption{The first section gives an overview of related work. Section two compares a content-based baseline to the hybrid approach. Section three shows the outcome of experiments with a larger text encoder (bert-large). The last section gives results for models trained on AudioCaps and ClothoV2. Values with $<$ are upper bounds estimated from reported recall values.}
\label{tab:main_results}
\begin{tabular}{llll|lllll|lllll}
\hline
\multicolumn{4}{c|}{model} & \multicolumn{5}{c|}{ClothoV2} & \multicolumn{5}{c}{AudioCaps} \\ \hline
\begin{tabular}[c]{@{}l@{}}name /\\ variation\end{tabular} & \begin{tabular}[c]{@{}l@{}}extra \\ data\end{tabular} & \begin{tabular}[c]{@{}l@{}}meta\\ data\end{tabular} & fusion & map@10 & \begin{tabular}[c]{@{}l@{}}$\Delta$ map\\ @10\end{tabular} & R@1 & R@5 & R@10 & map@10 & \begin{tabular}[c]{@{}l@{}}$\Delta$ map\\ @10\end{tabular} & R@1 & R@5 & R@10 \\ \hline
WavCaps \cite{wavcaps} & \cmark & \xmark & \xmark & $<35.97$ &  & 21.2 & 46.4 & 59.4 & $<54.13$ &  & 34.7 & 69.1 & 82.5 \\
Cacophony \cite{cacophony} & \cmark & \xmark & \xmark & $<35.20$ &  & 20.2 & 45.9 & 58.8 & $<60.00$ &  & 41.0 & 75.3 & 86.4 \\
DCASE23 \cite{DCASE2023} & \cmark & \xmark & \xmark & 36.65 &  & 24.26 & 53.89 & 66.87 &  &  &  &  &  \\ \hline
baseline & \xmark & \xmark & \xmark & 26.8 & $\pm$ 0 & 15.81 & 41.31 & 56.17 & 54.14 & $\pm$ 0 & 39.04 & 75.04 & 87.50 \\
hybrid & \xmark & OS/ CS & mid & 28.25 & +1.45 & 16.98 & 43.53 & 57.42 & 57.4 & +3.26 & 42.11 & 78.50 & 89.53 \\
hybrid & \xmark & OS/ CS & late & 29.16 & +2.36 & 18.17 & 43.46 & 56.93 & 57.83 & +3.69 & 42.75 & 78.65 & 89.77 \\
hybrid & \xmark & FS & mid & 33.57 & +6.77 & 22.37 & 48.50 & 61.80 &  &  &  &  &  \\
hybrid & \xmark & FS & late & 35.62 & +8.82 & 24.87 & 50.05 & 62.36 &  &  &  &  &  \\ \hline
large-baseline & \xmark & \xmark & \xmark & 27.98 & $\pm$ 0 & 16.68 & 43.32 & 57.85 & 55.41 & $\pm$ 0 & 40.15 & 76.50 & 88.18 \\
large-hybrid & \xmark & OS/ CS & late & 29.88 & +1.9 & 18.39 & 45.05 & 58.62 & 58.56 & +3.15 & 43.47 & 79.38 & 90.16 \\
large-hybrid & \xmark & FS & late & 37.01 & +9.03 & 26.2 & 51.14 & 63.95 &  &  &  &  &  \\ \hline
baseline & \cmark & \xmark & \xmark & 29.94 & $\pm$ 0 & 18.74 & 45.06 & 58.95 & 54.00 & $\pm$ 0 & 38.75 & 75.01 & 87.46 \\
hybrid & \cmark & OS/ CS & late & 31.48 & +1.54 & 20.08 & 46.77 & 60.41 & 57.82 & +3.82 & 42.79 & 78.48 & 89.86 \\ \hline
\end{tabular}
\end{table*}

Table \ref{tab:main_results} gives an overview of the performance of a variety of retrieval models. The first (top) section refers to systems from related work. We note that our content-based baselines trained exclusively on AudioCaps or Clotho are weaker because they are trained on less data. If trained on both datasets (last section of the Table \ref{tab:main_results}), the baseline becomes more competitive with the state-of-the-art systems. The remaining sections in Table \ref{tab:main_results} compare the performance of the pure content-based baseline to systems using different metadata types and modality fusion approaches. We discuss the results in greater detail below.

\subsection{Does the use of metadata lead to improved retrieval performance compared to a pure content-based approach?} The results in the second section of Table \ref{tab:main_results} suggest that including any of the three investigated metadata types to represent the item in the retrieval space leads to an improvement over the pure content-based baseline.

For ClothoV2, we observed a $2.36$ pp. improvement when using the OS tags as metadata. When using the FS descriptions as meta-data, we observed an even greater improvement of $8.82$ pp. map@10. However, the latter result must be interpreted with caution because of the potential positive bias that could arise from the high similarity between the FS metadata and the natural-language queries. In fact, some of the captions differ only in a few words and an untrained retrieval model initialized with pre-trained modality encoders achieves a R@1 of $8.96$. For practical applications, the retrieval performance will depend strongly on the similarity between query and description. Under less ideal conditions, we expect the actual improvement to be lower. However, we hypothesize that multi-sentence, unfiltered texts can still improve retrieval performance.
For the AudioCaps benchmark, we observed an improvement of $3.69$ pp. mAP@10 when using the CS tags as metadata.

We further repeated our experiments with a larger text embedding model (bert-large) to strengthen the validity of our results. The outcomes are presented in the third section of Table \ref{tab:main_results}. We observed comparable improvements over the baseline as we did for the smaller text embedding model. This indicates that the hybrid approach is general enough to be combined with other advancements, such as larger pre-trained modality encoders or new training objectives.

\subsection{Does hybrid retrieval benefit from modeling crossmodal interactions between audio and metadata embeddings?}
Section two of Table \ref{tab:main_results} compares models with late and mid-level fusion of audio and metadata. While the performance improved over the content-based baseline for both methods, we note that the late fusion approach tends to be better at ranking the items in general. This is surprising, but previous work \cite{timing} has suggested that retrieval models mostly focus on nouns and verbs, so matching the keywords and the descriptions directly is probably less error-prone. A modality fusion approach that is not based on the MMT architecture could potentially lead to more competitive results.

\subsection{How does combining open- and closed-set tags impact the performance?}
 AudioCaps and Clotho are often joined for training to increase the number of item-query pairs and boost retrieval performance. We are interested in using a similar approach for the hybrid architecture, which raises the question of whether different sources of metadata (OS and CS tags) can be combined for training and if this leads to similar improvements. Results are given in the third section of Table \ref{tab:main_results}. Training on this combined dataset improved the results on the ClothoV2 benchmark by $3.14$ and $2.32$ pp. map@10 for the content-based and hybrid approach, respectively. On the AudioCaps benchmark, the performance decreased slightly for the baseline and the hybrid model. This discrepancy could be attributed to the different characteristics of the data sets, such as word frequencies and audio length. Despite this, the hybrid method still improves the map@10 by $3.82$ pp. on the AudioCaps benchmark.

 \subsection{How do artificial captions generated from metadata impact hybrid models?}
Generating artificial audio captions from metadata with large language models \cite{wavcaps, cacophony, DCASE2023} has become a popular strategy because it is cheaper than hiring human annotators. We're interested in whether these artificial captions can be exploited for the hybrid approach as well. 
To this end, we train a content-based and a hybrid model on the FreeSound subset of WavCaps \cite{wavcaps}. For the hybrid model, we use the open set keywords as metadata; those keywords were also used to generate the artificial captions. The results are given in Table \ref{table:wavcaps_results}. We observe a notable drop in performance for the hybrid approach. It is likely that this is because the hybrid retrieval model focuses mostly on the high similarity between metadata and the keywords and neglects the audio signal during training. This high metadata caption similarity is not present in the test set, and consequently, the retrieval performance deteriorates.

\begin{table}[t]
\centering
\caption{ClothoV2 benchmark results when trained on the FreeSound subset of WavCaps. Captions of WavCaps were generated from metadata.}
\label{table:wavcaps_results}
\begin{tabular}{@{}lllll@{}}
\toprule
metadata & map@10 & R@1   & R@5   & R@10  \\ \midrule
none & \textbf{30.35}  & \textbf{18.74} & \textbf{45.87} & \textbf{59.61} \\
tags  & 27.67  & 17.21 & 41.76 & 54.94 \\ \bottomrule
\end{tabular}
\end{table}


\subsection{Are there benefits in sharing the text embedding model?}
Our hybrid architecture shares the text encoder for query and metadata embedding. We validate this choice by retraining the late fusion model with separate text encoders for query and metadata. The models with separate and shared text encoders achieved 28.05 and 29.16 mAP@10, respectively, which indicates that the hybrid approach benefits from parameter sharing.




\section{Conclusion}

This study investigated a hybrid metadata and content-based approach for language-based audio retrieval. We identified both open and closed-set keywords and natural language descriptions as suitable candidates to improve retrieval performance. Future work on hybrid retrieval models should also consider noisier keywords, unconstrained full-sentence descriptions, and missing metadata to mimic more realistic conditions. A comparison of two feature fusion approaches, one based on conceptually simple late fusion and the other on the multimodal transformer architecture, showed that both versions led to improvements over the content-based baseline. Surprisingly, the simpler late fusion strategy yielded slightly superior results. A more in-depth investigation of fusion methods would be needed to identify if this is a general trend or if it can be addressed to the MMT architecture. We further found that this hybrid approach does not pair well with captions that were generated from metadata, presumably because the model learns to rely on a high caption-metadata similarity, which is not present in the testing data.

\section*{Acknowledgment}
The LIT AI Lab is financed by the Federal State of Upper Austria. The computational results presented in this work have been partially achieved using the Vienna Scientific Cluster.


\bibliographystyle{IEEEtran}
\bibliography{conference_101719}

\end{document}